\documentclass[12pt]{article}
\def\be{\begin{equation}}
\def\ee{\end{equation}}

\begin{document}

\hsize=6in
\vsize=9in

\begin{center}
{\Large\bf Measuring the deviation from maximal mixing of atmospheric
muon neutrinos at INO} \\[.2in]
{\it Talk given at the PANIC 05 conference, Santa Fe, Oct 24-28, 2005}\\[.2in]
Probir Roy 
\end{center}
\begin{flushleft}
\hspace*{4.5cm} Tata Institute of Fundamental Research \\
\hspace*{4.5cm} Homi Bhabha Road, Mumbai 400 005, India \\
\hspace*{4.5cm} probir@theory.tifr.res.in
\end{flushleft}

{\bf
\begin{enumerate}
\item[{$\bullet$}] Neutrino factfile and $D \equiv 1/2 -
\sin^2\theta_{23}$
\vspace*{-.2cm}
\item[{$\bullet$}] $\nu_\mu,\bar\nu_\mu$ survival probabilities and
binned up-down asymmetries 
\vspace*{-.2cm}
\item[{$\bullet$}] Error analysis and sensitivity to $D$
\end{enumerate}
}

\noindent This talk is based on my work [1] with S. Choubey.
\bigskip

\noindent {\bf $\bullet$ Neutrino factfile}
\medskip

We take a CPT-conserving framework with only three active light
neutrinos: $\nu_{e,\mu,\tau}$ as flavor eigenstates and
$\nu_{1,2,3}$ as mass eigenstates with real, positive mass eigenvalues
$m_{1,2,3}$ and use $\Delta_{ij} \equiv m_i^2 - m^2_j$.  We relate
the two sets of eigenstates by
\be
\left(\matrix{\nu_e \cr \nu_\mu \cr \nu_\tau}\right) = U_\nu
\left(\matrix{\nu_1 \cr \nu_2 \cr \nu_3}\right), \ \ \ U_\nu \equiv
U_{PMNS} (\theta_{12},\theta_{23},\theta_{13},\delta).
\label{one}
\ee
Here $U_{PMNS}$ is defined in the standard CKM parametrization with 
possible Majorana phases ignored.  We define
\be
D \equiv 1/2 - \sin^2 \theta_{23}.
\label{two}
\ee
\[
\begin{tabular}{|c|}
\hline \\
$0.0085$ eV $< \sqrt{m^2_2 - m^2_1} < 0.0094$ eV \\[.1cm]
$0.041$ eV $< \sqrt{|m^2_3 - m^2_2|} < 0.057$ eV \\[.1cm]
$30^\circ < \theta_{12} < 38^\circ$ \\[.1cm] 
$36^\circ < \theta_{23} < 54^\circ$, i.e. $-0.16 < D < 0.16$ \\[.1cm]
$\theta_{13} < 12^\circ$ \\[.1cm]
$\delta =$ ? \\
\hline
\end{tabular} 
\] \begin{center}
Table 1. Known $3\sigma$ limits on neutrino parameters \end{center}

\noindent The known $3\sigma$ limits on the relevant neutrino mass and
mixing parameters 
are given in Table 1.  It is also instructive to keep in mind that
$-0.10 < D < 0.10$ at $2\sigma$ and $-0.07 < D < 0.07$ at $1\sigma$ levels.

\newpage

\noindent {\bf $\bullet$ $\nu_\mu,\bar\nu_\mu$ survival probabilities
and binned up-down asymmetries}
\medskip

We explore the measurability of $D$ from atmospheric muon neutrino and
antineutrino studies in a large magnetized iron calorimeter like ICAL
-- as proposed [2] in the India-based Neutrino Observatory INO.
Effects due to matter and subleading oscillations on account of the
solar neutrino mass scale $\Delta_{21}$ cause departures from the
simple $\sin^22\theta_{23}$ dependence of 2-flavor vacuum oscillations
in the survival probability $P_{\mu\mu}$ $(P_{\bar\mu\bar\mu})$ for a
$\nu_\mu$ $(\bar\nu_\mu)$.  Both these effects as well as vacuum
oscillation terms play a role in the measurement of $|D|$.  However,
matter effects are found to be crucial in resolving the octant
ambiguity $\sin^2\theta_{23} \leftrightarrow 1 - \sin^2\theta_{23}$,
i.e. whether $\theta_{23}$ is greater or less than $\pi/4$, through a
determination of the sign of $D$.  Our numerical code exactly solves the three
generation neutrino equations of motion in earth matter with the PREM
density profile.  We assume normal neutrino mass
ordering, but we can do the analysis for the inverted case.  We fix some of
the oscillation parameters at chosen benchmark values, to wit
$\Delta_{31} = 2 \times 10^{-3} \ {\rm eV}^2$, $\Delta_{21} = 8 \times
10^{-5} \ {\rm eV}^2$, $\sin^2\theta_{12} = 0.28$, $\sin^2
2\theta_{13} = 0.1$, $\delta = 0$. Using the atmospheric neutrino flux
of Honda et al [3] and the DIS 
cross sections of the CTEQ collaboration [4], we have simulated 14000
(7000) $\nu_\mu$ $(\bar\nu_\mu)$ events for 1 MtonY of exposure for
ICAL assuming a 50\% trigger efficiency.  We have binned our data {\it
both} in energy $E$ and the cosine 
of the zenith angle $\xi$.  The difference between the $\nu_\mu$ and
$\bar\nu_\mu$ up-down asymmetry
ratios $U_N/D_N - U_A/D_A$ is shown in Fig. 1 with solid (dashed) pink and black lines corresponding to $\sin^2
\theta_{23} = 0.36$ and $0.5$ respectively in matter (vacuum).
The largest usable effects are in the topmost right panel (the 
lowermost right panel having problems due to earth-core
uncertainties) with the 5-7 GeV energy bin corresponding to a
maximum SPMAX in $P_{\mu\mu}$.  

\vspace*{2.25in}
\begin{figure}[h]
\includegraphics{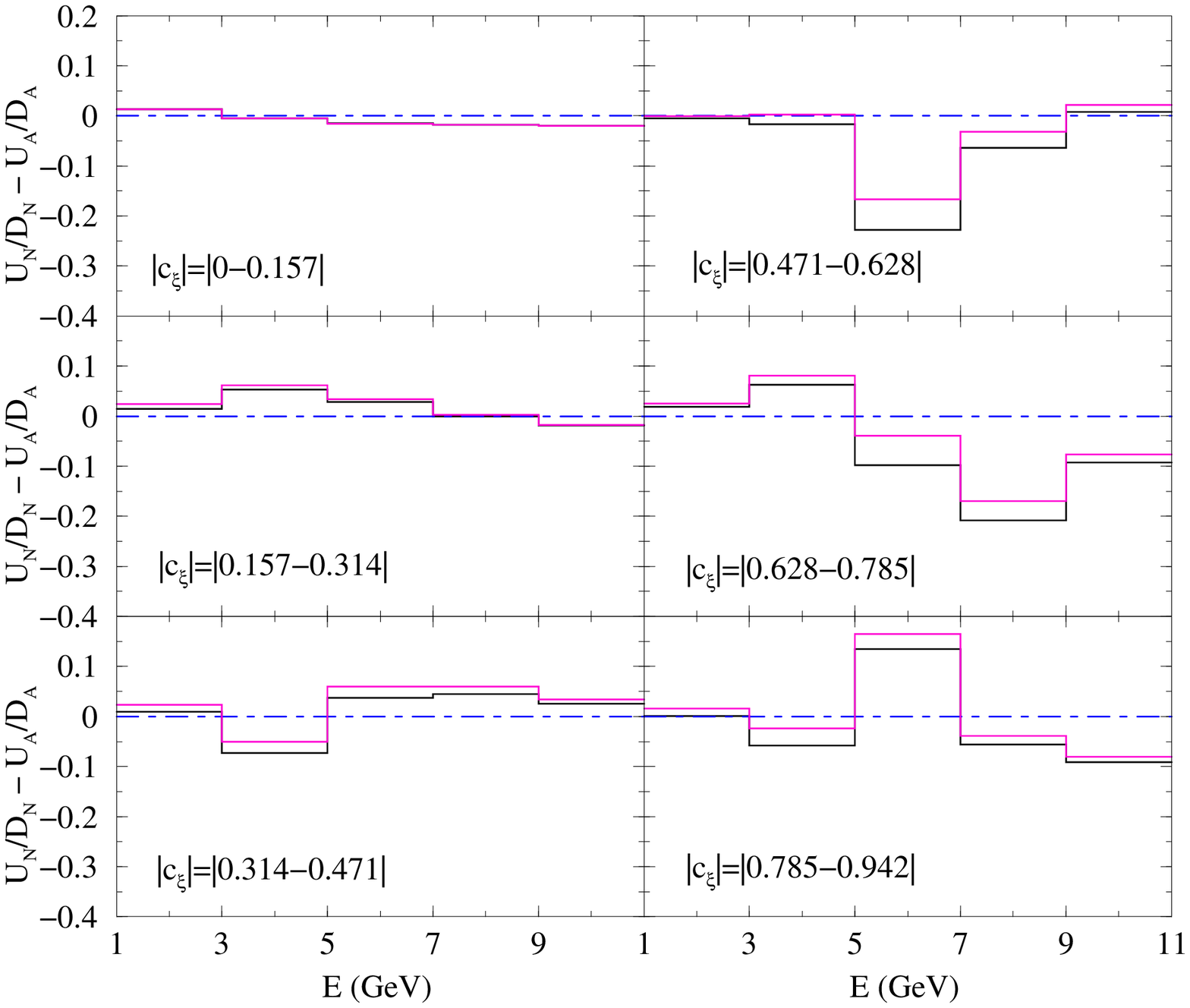} 
\label{fig:panictalk_fig1.eps}
\end{figure}
\begin{center} Fig. 1. The difference $U_N/D_N - U_A/D_A$ for various
bins of $E$ and $|c_\xi|$. \end{center}

\newpage

\noindent {\bf $\bullet$ Error analysis and sensitivity to $D$}
\medskip

We have done a $\chi^2$ analysis using the pull approach, as detailed
in [1].  Fig. 2 shows the regions of $\sin^2\theta_{13}$ for which a
maximal $\theta_{23}$ can be rejected at $3\sigma$ (yellow), $2\sigma$
(green) and $1\sigma$ (magenta) levels.  The capability to resolve the
octant ambiguity   

\vspace*{1.5in}
\begin{figure}[h]
\includegraphics{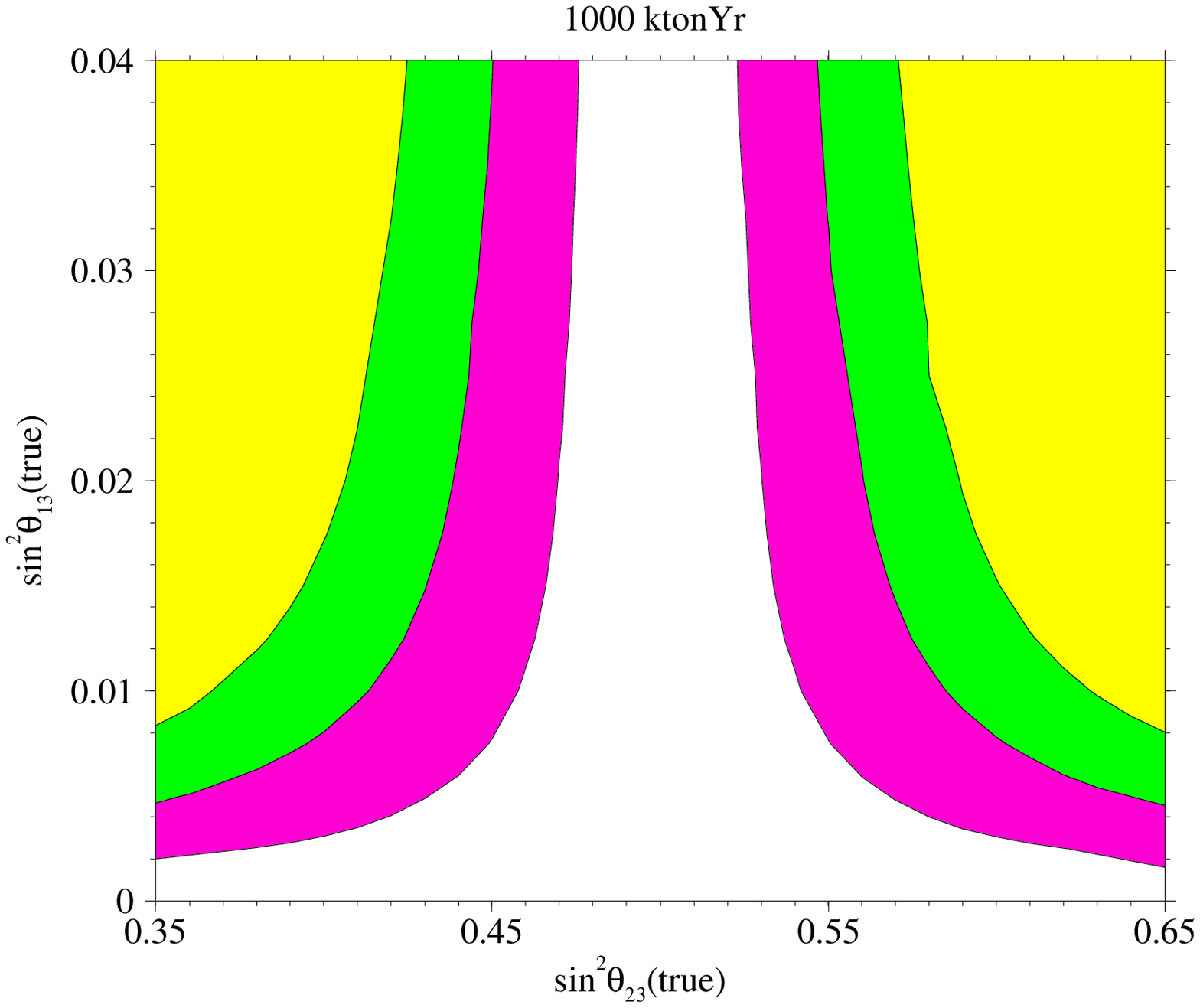} 
\label{fig:panictalk_fig2.eps}
\end{figure}
\begin{center} Fig. 2. Sensitivity to deviation from maximal
$\theta_{23}$ in terms of $\theta_{13}$ \end{center}

\noindent in $\theta_{23}$ is shown in the $\Delta \chi^2$
vs. $\sin^2\theta_{23}$ 
plot of Fig. 3 where $\Delta \chi^2$ is the difference in $\chi^2$ for
`true octant' and `false octant' values.  The latter is made possible by the
unique capability of a large magnetized iron calorimeter to isolate
matter effects by distinguishing between $\nu_\mu$ and $\bar\nu_\mu$
events. 

\vspace*{1.9in}
\begin{figure}[h]
\includegraphics{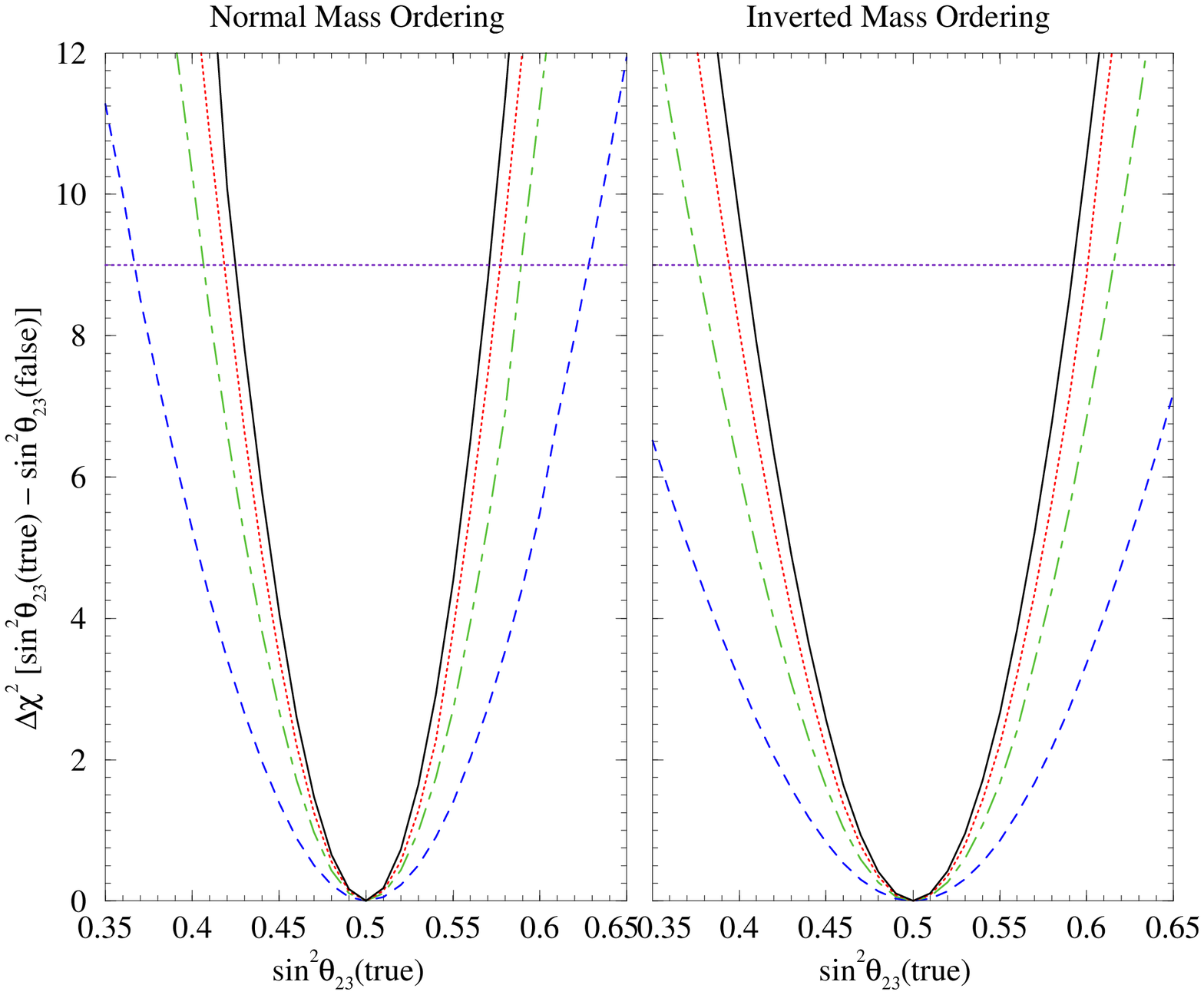} 
\label{fig:panictalk_fig3.eps}
\end{figure}
\begin{center} Fig. 3. Capability of octant ambiguity resolution
\end{center}

\begin{enumerate}
\item[{[1]}] S. Choubey and P. Roy, hep-ph/0509197, to appear in
Phys. Rev. D.
\item[{[2]}]
http:$\backslash\backslash$www.imsc.res.in$\backslash\sim$~ino$\backslash$. 
\item[{[3]}] M. Honda et al., Phys. Rev. D{\bf 70}, 043008 (2004).
\item[{[4]}]
http:$\backslash\backslash$user.pa.msu.edu$\backslash$wkt$\backslash$cteq6$\backslash$cteq6pdf.html
\end{enumerate}

\end{document}